\documentclass[aps,prl,twocolumn,showpacs]{revtex4}

\usepackage[dvips]{graphicx}
\begin{document}
\title{Voltage-Driven Spin-Transfer Torque in a Magnetic Particle}
\author{P. Gartland}
 \email{pgartland3@gatech.edu}
 \affiliation{School of Physics, Georgia Institute of Technology, 837 State Street, Atlanta, Georgia 30332, USA}
 \author{D. Davidovi\'c}%
 \email{dragomir.davidovic@physics.gatech.edu.}
\affiliation{School of Physics, Georgia Institute of Technology, 837 State Street, Atlanta, Georgia 30332, USA}

\date{\today}


\begin{abstract}

We discuss a spin-transfer torque device, where the role of the soft ferromagnetic layer is played by a magnetic particle or a magnetic molecule, in weak tunnel contact with two spin polarized leads. We investigate if the magnetization of the particle can be  manipulated electronically, in the regime where the critical current for magnetization switching is negligibly weak, which could be due to the reduced particle dimensions. Using master equation simulations to evaluate the effects of spin-orbit anisotropy energy fluctuations on spin-transfer, we obtain reliable reading and writing of the magnetization state of such magnetic particle, and find that the device relies on a critical voltage rather than a critical current. The critical voltage is governed by the spin-orbit energy shifts of discrete levels in the particle.
This finding opens a possibility to significantly reduce the power dissipation involved in spin-transfer torque switching, by using very small magnetic particles or molecules.
%
\end{abstract}

\maketitle
In recent years, the miniaturization of magnets has approached the scale of single molecules~\cite{bogani,heersche,miyamachi,moon,osorio,osorio2,sanvito}. While there are many measurement techniques for determining the magnetic state of such molecules, electron transport is particularly important for integrating the molecules into a microelectronic system~\cite{aradhya}. The reduced dimensions of magnetic molecules pose both challenges and advantages. A primary challenge is due to the fact that the energy barrier for magnetic switching ($E_B$) is suppressed in proportion with the volume of the magnet, which weakens the directional stability of the magnetization subject to thermal or electronic perturbations. However, the weakened barrier could also be viewed as an advantage if the external perturbations are properly controlled to manipulate magnetic switching. Consider spin-transfer torque (STT) switching in a metallic nanomagnet~\cite{slonczewski1,berger,ralph2,katine1}. The switching is usually achieved by applying a spin-polarized current through the nanomagnet, via Ohmic contacts between a ferromagnetic lead and the nanomagnet. STT-switching is normally a current-driven effect.
Using the expression for current found in reference \onlinecite{slonczewski1}, it can be shown that the critical current for magnetization switching due to  STT is proportional to $e\alpha E_B/\eta\hbar$, where $e$ is the electronic charge, $\alpha$ is the Gilbert damping parameter, and $\eta$ is the efficiency ratio dependent on both the spin polarization in the leads $P$ and the angle between the equilibrium magnetization in the lead and the nanomagnet~\cite{slonczewski1}. In larger nanomagnets, $E_B$ is large and the resulting critical current can be associated with large power dissipation, which is a well known problem for applications. By reducing the size to the molecular scale, $E_B$ can be significantly reduced, leading to the possibility of much lower critical current. Alternatively, a possible reduction in the critical current could be achieved with an increased spin relaxation time~\cite{deshmukh,yakushiji2,jiang}, which could reduce $\alpha$. In this article, we consider a magnetic particle or a magnetic molecule, making weak tunneling contact between two ferromagnetic leads, and assume a vanishingly small critical current for STT switching. We address the question if the magnetization direction in such a regime can be reliably measured and manipulated, and find a voltage driven mechanism that controls STT-switching. In that regime, the spin-transfer is dominated by fluctuations of spin-orbit anisotropy energy, and predictable magnetic switching can be induced by applying a critical voltage, independent of the size of the tunneling current.  By applying a voltage smaller than the critical voltage, the magnetization direction can be read noninvasively, without inducing magnetic switching.

\begin{figure}[ht!]
\centering
\includegraphics[width=0.95\columnwidth]{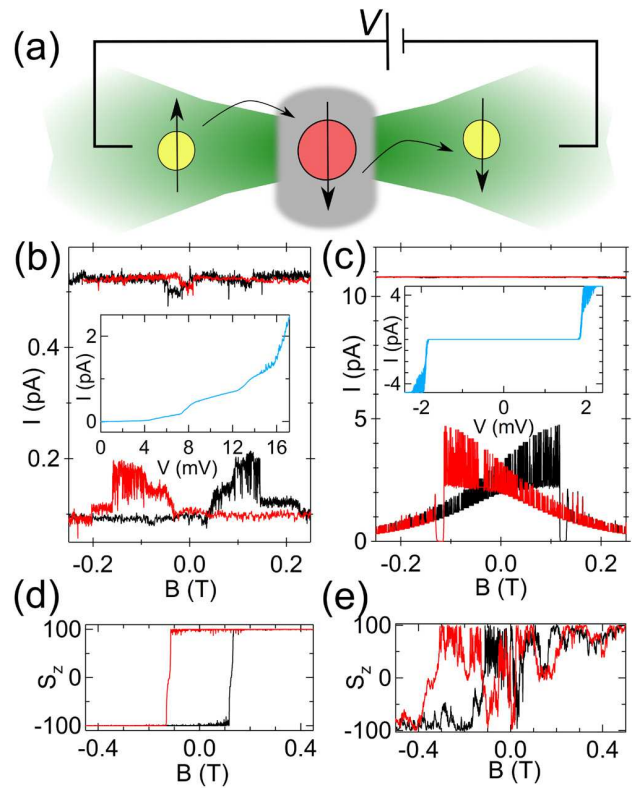}
\caption{(a) Junction geometry with either no spin polarization ($P = 0$) in leads, and an applied magnetic field ($B\neq 0$), or no external magnetic field ($B=0$), and non-zero spin polarization in leads ($P\neq 0$). Red circle corresponds to magnetic particle with net magnetization in direction of black arrow. Bias voltage $V$ is applied on the left lead, relative to the right lead. (b) Experimental hysteresis loop data at low and high bias voltage values at $T=350$ mK. Inset is the sample $I(V)$ curve. (c) Simulated current hysteresis loops at low ($V=1.9$ mV) and high ($V=2.4$ mV) bias, with inset simulated $I(V)$ curve for $P=0$ and current onset threshold of $V_{se}=\pm 1.8$mV. Top curve is offset vertically by 6 pA for clarity. (d) Simulated hysteresis of particle magnetization, corresponding to low bias data in (c). (e) Simulated non-hysteretic switching, corresponding to high bias data in (c).  For all hysteresis loops, black (red) corresponds to field sweep in positive (negative) direction.}
\label{spjfig1}
\end{figure}

Recently, we have demonstrated experimentally that single Ni particles 2 to 3 nm in diameter, embedded in double-tunnel junctions, exhibit hysteresis based on the applied bias voltage~\cite{gartland2}. A schematic of the device that we studied is shown in figure \ref{spjfig1}(a). At low temperature, the particle exhibits Coulomb blockade at low bias voltages, and sequential electron tunneling at higher bias voltages. The presence of hysteresis was found to be governed primarily by the voltage applied across the junction, rather than being controlled by the tunneling current.  However, our prior experimental work involved the coupling of the Ni particle to normal metal (Al) leads, which lack spin-polarization. Crucial to the voltage control of magnetic hysteresis in nanomagnets is the presence of spin-orbit energy shifts $\epsilon_{SO}$ of the discrete energy levels of the particle, which vary with the direction of the magnetization\cite{deshmukh,cehovin}.
The voltage-control results from an effective magnetization blockade\cite{gartland2}, which arises from electron tunneling transitions with energy
that increases as the particle magnetization is displaced from the easy axis. The value of the magnetization blockade energy is given by the tunneling transition energy $\Delta E$, for which $d\Delta E/d S_z=0$, and the magnetization is closest to the easy axis. Here, $S_z$ is the particle spin component along the easy axis. Finding this transition energy requires diagonalization of  the particle's magnetic Hamiltonians, and in a  typical case, we find the magnetization blockade energy to be $\sim 0.65\epsilon_{SO}$ \cite{gartland2}. At low bias voltage, magnetization blockade prohibits electron tunneling transitions that would perturb the magnetization beyond a certain angle from its easy axis. At bias voltages larger than $\epsilon_{SO}/e$ (relative to the Coulomb blockade threshold), the magnetization blockade is surmounted. The result is a voltage-controlled magnetic hysteresis over a bias range on the order of $\epsilon_{SO}/e$, governed by the shifts in spin-orbit anisotropy energy.

Figure \ref{spjfig1}(a) displays a generalization of such a configuration, in which we propose a non-zero spin polarization in the source and drain leads. In figure \ref{spjfig1}(a), the grey region corresponds to a double-tunnel barrier, and the red circle corresponds to the single-domain magnetic particle. The straight black arrows in this figure and in subsequent figures correspond to the predominant spin polarization (in the $+z$ or $-z$ direction for $\uparrow$ or $\downarrow$, respectively). The collection of spins in the magnetic particle determine the direction of the magnetization.

In this work, we model the effects of single electron tunneling by use of master equations in the same procedure used in our previous work \cite{gartland2}. In doing so, we explore the viability of such an experimental realization as is displayed in Figure \ref{spjfig1} (a).
The magnetic Hamiltonian under consideration is given by the following two alternating operators. For the N electron particle, $H_0 = -KS_z^2/S_0+2\mu_B BS_z$,
where $K$ is a coefficient for the uniaxial anisotropy of the particle, $B$ is the applied magnetic field, and $S_z$ is the spin operator in
the $z-$direction, $\mu_{B}$ is the Bohr magneton, and $S_0$ is the ground state spin of the particle in units of $\hbar$.
For the N+1 electron particle, $H_1= H_0+\epsilon\left[\cos{\theta_{SE}}S_z+\sin{\theta_{SE}}S_x\right]^2/S_0^2+\epsilon_zS_z^2/S_0^2+E_0$.
In the latter case, we include, in addition to the N electron case, the terms $\epsilon$ and $\epsilon_z$ resulting from the spin-orbit energy shifts.
$\theta_{SE}$ is the angle of the new anisotropy term arising from the additional electron, and $E_0$ is a constant offset term that depends on the
Coulomb blockade and discrete electron-hole quasiparticle spacing~\cite{Note1}. 
We neglect coupling between the magnetization and the thermal bath. %
In the previous experimental and computational models\cite{gartland2}, as is shown in figure \ref{spjfig1}(b)-(f) a magnetic field is swept rather than using spin-polarized leads. The measured current hysteresis loops as a function of magnetic field are displayed in figure \ref{spjfig1}(b). The top (bottom) curves correspond to the current response at high (low) voltages relative to the magnetization blockade voltage. The dip in current prior to the zero field crossing is an artifact due to the superconducting magnet. The inset shows the tunneling current as a function of applied bias voltage. A typical stochastic realization of he simulated current hysteresis loops at low and high bias relative to the magnetization blockade are displayed in \ref{spjfig1}(c), along with a simulated $I(V)$ curve in the inset. The simulated magnetization hysteresis loops that correspond to the current loops in figure \ref{spjfig1}(c) are displayed in figure \ref{spjfig1}(d) and (e), for low and high voltages, respectively, relative to $\epsilon_{SO}/e$ above the sequential electron tunneling threshold $V_{se}$. The experimental data and the simulations demonstrate robust magnetic hysteresis at low voltage and random magnetic switching at high bias voltage. The characteristic voltage scale that differentiates between the two regimes corresponds to the magnetization blockade energy.


\begin{figure}[ht!]
\centering
\includegraphics[width=0.9\columnwidth]{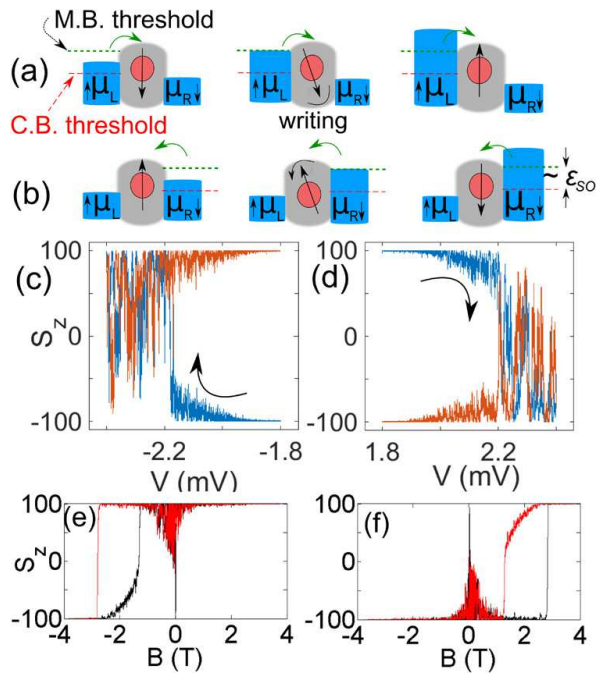}
\caption{Writing the magnetic state with spin-polarized leads $P_L = 0.5, P_R=-0.5$. (a) Illustration of the electrochemical potential of the left lead ($\mu_L$) for the forward writing process ($M_{\downarrow}\rightarrow M_{\uparrow}$), implying a negative $V$. Long red dashed line is the coulomb blockade threshold for sequential electron tunneling. When the writing threshold, magnetization blockade (smaller green dotted line) is reached, the particle magnetization flips directions. (b) Illustration of reverse state writing process ($M_{\uparrow}\rightarrow M_{\downarrow}$). Green arrows indicate electron tunneling direction. (c) Magnetization vs voltage during forward magnetic state writing process illustrated in (a). (d) Magnetization vs voltage for reverse magnetic state writing process as illustrated in (b). In both (c) and (d), blue (orange) correspond to magnetization during positive (negative) magnitude ramp of bias. (e) Magnetic field hysteresis loop with $P_L=0.5$, $P_R=-0.5$ at $V=-2.2$ mV. Black (red) corresponds to field sweep in positive (negative) direction. (f) same as (e), but with $P_L=-0.5$, $P_R=0.5$.}
\label{spjfig2}
\end{figure}

Figure \ref{spjfig2}(a) and (b) illustrate the Coulomb blockade threshold and the magnetization blockade threshold. When the electrochemical potential of the lead is raised above the first blockade, sequential electron transport is initiated as indicated by the curved green arrows. When the electrochemical potential is increased above the magnetization blockade threshold, the spin-polarized leads initiate the particle magnetization state writing process. In (a), the particle is initially in the $M_{\downarrow}$ state. The voltage is swept linear ramp from $-1.8$ mV to $-2.4$ mV, and back to $-1.8$ mV, as is shown in \ref{spjfig2}(c). When the electrochemical potential in the left lead reaches the writing threshold, indicated by the small dotted line in (a), the magnetization of the particle flips into the $M_{\uparrow}$ state. The voltage threshold for sequential electron tunneling is $V_{se}=\pm 1.8$ mV, while the voltage required for flipping the magnetic state is approximately $V_{w}=\pm 2.2$ mV.
Figure \ref{spjfig2}(d) displays the magnetization during the reverse writing process.
Consider the forward writing process (that is, using a negative bias voltage to write the $M_{\uparrow}$ state). As the voltage magnitude rises between $V_{se}$ and $V_{w}$, the magnetization (given by $S_z$) begins to fluctuate about its energetic minimum of $S_z=-100$. When the writing threshold potential is reached, the magnetization blockade is surmounted, and the magnetization flips as indicated by the sudden jump of $S_z$ around $V=-2.2$meV. Because the applied bias is still large at this point, the magnetization continues to fluctuate about its other energetic minimum state of $S_z=+100$. When the potential is reduced to its initial value, the fluctuations diminish as the magnetization relaxes into the $M_{\uparrow}$ state.

Similarly, the reverse writing process is displayed in figure \ref{spjfig2}(b) and (d), wherein a positive bias voltage greater than $2.2$ mV induces a switch of the particle into the $M_{\downarrow}$ state. The same result is obtained as in the forward writing process in figure \ref{spjfig2}(a) and (c). Thus, the sign of voltage can be used to write the binary state of the magnetic particle. In figure \ref{spjfig2}(e) and (f), we simulate hysteresis loops as a function of magnetic field for spin-polarized leads held at a bias voltage of -2.2 mV, the magnitude of which is above the magnetization blockade threshold of -2.16 mV. As a result, we observe an effective exchange bias due to spin accumulation on the particle. If the magnitude of the bias voltage is below the magnetization blockade threshold, the hysteresis loop appears qualitatively the same as in the $P=0$ case, as the blockade protects the particle from switching near zero field.
In figure \ref{spjfig2}(e), the left and right leads have spin polarization values of $P_L=0.5$ and $P_R=-0.5$, respectively. In figure \ref{spjfig2}(f), $P_L = -0.5$ and $P_R = 0.5$. In both (e) and (f), the black (red) corresponds to a positive (negative) magnetic field sweep direction. In both of these sweeps, there is an increase in magnetization noise near zero field, as the particle has a small probability to flip into the opposite magnetization state. As we have observed in a previous work~\cite{jiang1}, the smaller magnetic spectrum spacing leads to the enhancement of spin-flip rate, due to the spin-orbit energy fluctuations. This is precisely the characteristic we require to allow STT switching in the proposed configuration. As the magnetic field increases, the spin-flip rate is reduced significantly when the Zeeman splitting energy approaches the spin-orbit energy shift $\epsilon_{SO}$.

\begin{figure}[ht!]
\centering
\includegraphics[width=0.9\columnwidth]{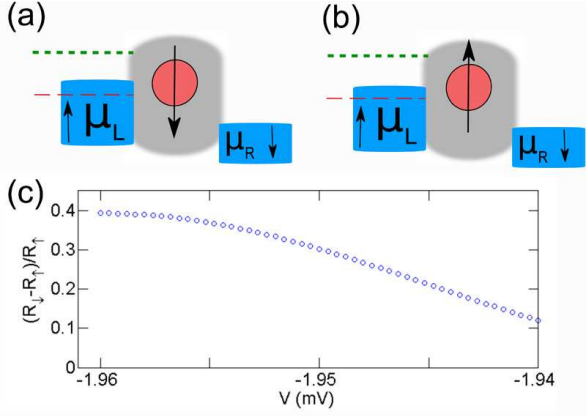}
\caption{Reading the magnetic state. (a) Sensing the $M_{\downarrow} $ state. (b) Sensing the $M_{\uparrow}$ state. (c) Ratio of the differential resistance measurements for $M_{\downarrow} $ : $M_{\uparrow}$. Electrochemical potential is always maintained well below the writing threshold, indicated by the small dotted green line in (a) and (b).}
\label{spjfig3}
\end{figure}

If the bias voltage is maintained well between $V_{se}$ and $V_{w}$, the differential resistance measurements can operate as a non-invasive sensor to determine the particle magnetization state.  Figure \ref{spjfig3}(a) and (b) illustrate the reading process for the $M_{\downarrow}$ and $M_{\uparrow}$ states, respectively. Figure \ref{spjfig3}(c) displays the differential resistance ($dV/dI$) ratios, averaged over time, for the $M_{\downarrow}$:$M_{\uparrow}$ states. As a function of bias, the ratio of differential conductance varies as much as $40\%$ for the different magnetization states. This results from the asymmetry in the tunneling resistance near the Coulomb blockade as a function of spin polarization mismatch.
As long as the electrochemical potential in the spin-polarized leads is maintained well below the writing threshold, the particle magnetization will only fluctuate weakly about its current energetic minimum orientation, allowing for the reproducible sensing of the magnetic state.

\begin{figure}[ht!]
\centering
\includegraphics[width=0.95\columnwidth]{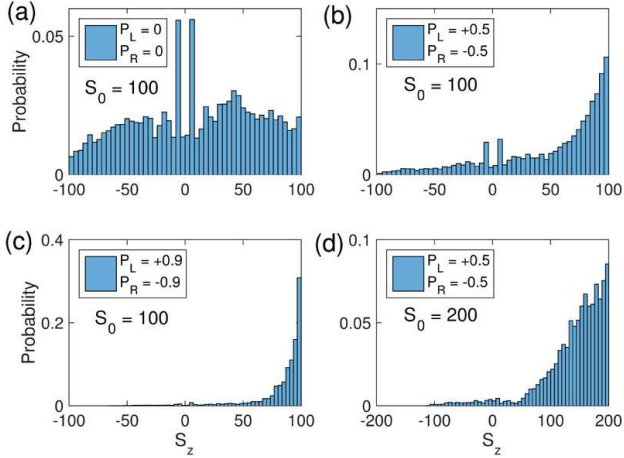}
\caption{Particle state histograms as a function of spin polarization in leads and particle size, at high bias voltages relative to $\epsilon_{SO}/e$. (a) Normal leads with no spin polarization. (b) Spin polarization of $\pm 0.5$ in left and right leads, respectively. (c) Spin polarization of $\pm 0.9$ in left and right leads, respectively. $S_0=100$ in (a),(b), and (c). (d) Spin polarization of $\pm 0.5$ in left and right leads, respectively, with larger particle size of $S_0=200$.}
\label{spjfig4}
\end{figure}

At high voltages compared with $V_{se}+\epsilon_{SO}/e$ during the particle writing process, there is a certain probability that the particle will switch back to its initial state before relaxing. Figure \ref{spjfig4} displays the histograms of the particle spin states, as a function of spin polarization in the leads and particle size.  In each of these cases, the bias was held at $V = -2.4$meV, which is above the magnetization blockade voltage, and would correspond to the forward state writing process as shown in figure \ref{spjfig2}(c). For each configuration, we estimate the reliability $r$ of successfully writing the state $M_{\uparrow}$ by taking the ratio of the sum of states with $S_z>0$ to the total sum of states. We chose such a definition because in the actual state writing process, we would reduce the magnitude of the bias voltage, and the particle would relax into whichever local minimum had the same sign ($S_z=\pm S_0$, based on the final sign of $S_z$ before lowering the voltage). For figure \ref{spjfig4}(a), the state histogram is evaluated for non-spin-polarized leads, as in the case for our previous experimental work~\cite{gartland2}. Not surprisingly, the non-polarized leads are ill-equipped to produce dependable switching of the particle state to $M_{\uparrow}$. However, as is shown in figure \ref{spjfig4}(b) and (c), the reliability of switching the particle becomes $83\%$ ($98\%$), with respective increases in spin polarization in the leads $P_L=0.5$ and $P_R=-0.5$ ($P_L=0.9$ and $P_R=-0.9$). A similar effect can be achieved by altering the size of the particle. For spin polarization in the leads of $P_L=0.5$ and $P_R=-0.5$, the reliability of switching a particle of spin $S_0=50$ ($S_0=200$) is found to be $78\%$ ($97\%$). The state histogram for the latter case of $S_0=200$ is displayed in figure \ref{spjfig4}(d). In the case of $P_L=0.25$ and $P_R=-0.25$ for $S_0=100$, we found a reliability of $75\%$. However, we can increase the effective reliability by use of the following procedure. First, we apply a voltage pulse of the desired sign to attempt writing a magnetic state. Then, we can apply a smaller bias voltage to read out the state. If the initial pulse has successfully written a state, then halt the procedure. If not, apply a second voltage pulse and check the success of the second attempted writing procedure. Assuming independent writing events, each dictated by a reliability $r$, with $0\le r \le 1$, the overall reliability of the repeated writing procedure will be $1-(1-r)^n$, where $n$ is the number of attempted writing events. If at any point between the writing attempts a successful state readout is achieved, then the procedure is stopped. Using such a scheme with $r=0.75$ and a maximum of 4 writing attempts, the new reliability of writing the desired state is raised above $99\%$.

In summary, we have presented a proposal for a generalized STT system in which the soft magnetic layer is composed of a magnetic molecule or a magnetic particle which exhibits an effective magnetization blockade due to spin-orbit shifts of discrete levels. Rather than relying on a critical current to induce magnetic switching, magnetic control in our proposed configuration is instead governed by the applied bias voltage from spin-polarized leads. Our simulations indicate that the proposed configuration is well suited for writing magnetic states with high repeatability, and for reading states in a non-invasive manner. This opens the possibility for a significant reduction in power dissipation in reduced scale STT devices.

This research was supported by the US Department of
Energy, Office of Basic Energy Sciences, Division of Materials Sciences and Engineering under Award DE-FG02-06ER46281.

\bibliography{career1}

\end{document}